\documentclass[english]{article}
\usepackage[affil-it]{authblk}
\usepackage{amsfonts} 
\usepackage[T1]{fontenc}
\usepackage[latin9]{inputenc}
\usepackage{refstyle}
\usepackage{color,hyperref}
    \catcode`\_=11\relax
    \newcommand\email[1]{\_email #1\q_nil}
    \def\_email#1@#2\q_nil{%
      \href{mailto:#1@#2}{{\emailfont #1\emailampersat #2}}
    }
    \newcommand\emailfont{\sffamily}
    \newcommand\emailampersat{{\color{red}\small@}}
    \catcode`\_=8\relax

\makeatletter

\RS@ifundefined{subref}
  {\def\RSsubtxt{section~}\newref{sub}{name = \RSsubtxt}}
  {}
\RS@ifundefined{thmref}
  {\def\RSthmtxt{theorem~}\newref{thm}{name = \RSthmtxt}}
  
  {}
\RS@ifundefined{lemref}
  {\def\RSlemtxt{lemma~}\newref{lem}{name = \RSlemtxt}}
  {}

\usepackage{amsmath}

\makeatother

\usepackage{babel}
\usepackage{mathrsfs}
\begin{document}
\title{Operational classical mechanics: Holonomic
Systems }
\author{A. D. Berm\'udez Manjarres}
\affil{\footnotesize Universidad Distrital Francisco Jos\'e de Caldas\\ Cra 7 No. 40B-53, Bogot\'a, Colombia\\ \email{ad.bermudez168@uniandes.edu.co}}
\maketitle
\maketitle
\begin{abstract}
We construct an operational formulation of classical mechanics without
presupposing previous results from analytical mechanics. In doing
so, we rediscover several results from analytical mechanics from an
entirely new perspective. We start by expressing the position and
velocity of point particles as the eigenvalues of self-adjoint operators
acting on a suitable Hilbert space. The concept of Holonomic constraint
is shown to be equivalent to a restriction to a linear subspace of
the free Hilbert space. The principal results we obtain are: (1) the
Lagrange equations of motion are derived without the use of D'Alembert
or Hamilton principles, (2) the constraining forces are obtained without
the use of Lagrange multipliers, (3) the passage from a position-velocity
to a position-momentum description of the movement is done without
the use of a Legendre transformation, (4) the Koopman-von Neumann
theory is obtained as a result of our ab initio operational approach,
(5) previous work on the Schwinger action principle for classical
systems is generalized to include holonomic constraints. 
\end{abstract}

\section{Introduction}

Inspired by the rise of quantum mechanics, Koopman\cite{KvN1} and
von Neumann \cite{KvN2} reformulated classical (Hamiltonian) statistical
mechanics in terms of operators acting on a Hilbert space (see \cite{mauro}
for a review). Since then, the main reasons to study the Koopman-von
Neumann (KvN) theory have been to investigate quantization rules and
to make other comparisons with the quantum theory \cite{compare1,compare2,compare3,compare4,compare5}.
Another active area of research involving the KvN theory is the development
of consistent quantum-classical hybrid theories \cite{hybrid0,hybrid1,hybrid2,hybrid3,hybrid4,hybrid5,hybrid6}.
A third and more recent line of research deals with quantum simulations
of non-linear systems \cite{algorithm1,algorithm2,algorithm3}.

However, apart from any connection with quantum physics, the operational
methods had produced results entirely within the classical domain
\cite{classical1,classical2,classical3,classical4,classical5,classical6,classical7,classical8,classical9,classical10}.
Hence, it seems worth considering the operational approach in a more
encompassing context than the KvN theory.

We will show in this work that it is possible to construct a classical
operational dynamics independently from previous results from analytical
mechanics (unlike the KvN theory that uses the Poisson brackets and
the Liouville theorem as the starting point). 

We will see that the operational approach gives new insights into
many old results of analytical mechanics. We will restrict ourselves
to holonomic scleronomous systems and leave the study of more general
systems for future works.

This work is organized as follows: in section 2 we review the operational
formulation of classical mechanics for the unconstrained case, as
developed in \cite{free1,free2}. Here we introduce most of the operators
and the ket basis for the Hilbert space we deal with in the rest of
the paper. As in quantum mechanics, the evolution of the system can
be given by a Schr\"{o}dinger equation for the state vectors, or by Heisenberg
equations for the operators. We show that the passage from a velocity
to a momentum description of the movement is given by a ``quantum''\footnote{We use \textquotedblleft quantum\textquotedblright{} throughout this
paper to indicate that the mathematical concept at hand originated
in the context of quantum mechanics, nevertheless, its use by us does
not imply any actual quantum physics.} canonical transformation as understood in \cite{QCT}.

In section 3 we define what we understand as a holonomic constraint
in our operational approach. We will see how our procedure naturally
leads to the concept of the tangent bundle of configuration space
as given in the geometrical expositions of analytical mechanics. 

In section 4 we will use the so-called ``quantum'' point transformation
\cite{dewitt,dewitt2} to give the operational version of generalized
coordinates and velocities. We emphasize that the change to generalized
coordinates is an example of a ``quantum'' canonical transformation.

Some of the most important results of this paper are in section 5
. We will see that our operational version of holonomic constraints
and the generalized coordinates allow us to derive the Lagrange equations
of motion without the use of D'Alembert or Hamilton principles. The
same procedure will allow us to obtain the constraining forces without
using Lagrange multipliers. We then perform a \textquotedblleft quantum\textquotedblright{}
canonical transformation to a momentum description of the movement
to get the KvN theory.

In section 6 we derive the Lagrange equation of motion from the variation
of the so-called Schwinger action, extending previous work on the
Schwinger action principle {[}38{]} to systems with holonomic constraints.
We comment on the possible relation of the Schwinger action principle
with the Gauss principle of least constraints.

\section{Unconstrained mechanics}

We review in this section the operational formulation of classical
mechanics for the unconstrained case. Several of the results will
be stated without proof, and we refer to {[}29{]} for the them and
more in-depth treatment.

Let us have a system of $N$ point particles that are free of constraints.
Classically, we have to know the position and velocity of each particle
to determine the state of this system, this is equivalent to specifying
a point in the tangent bundle of $\mathbb{R}^{3N}$, i.e., $T\mathbb{R}^{3N}\cong\mathbb{R}^{6N}$.
We will relax this definition of a classical state in favor of a statistical
description in terms of probability amplitudes. 

As a starting point of the operational formulation of classical dynamics,
we associate a ray $\left\{ \left|\mathbf{r}_{1},...,\mathbf{r}_{N};\mathbf{v}_{1},...,\mathbf{v}_{N}\right\rangle \right\} $
to each point in $T\mathbb{R}^{3N}$. The kets are postulated to obey
the orthonormality condition

\begin{equation}
\left\langle \mathbf{r}'_{1},...,\mathbf{r}'_{N};\mathbf{v}'_{1},...,\mathbf{v}'_{N}\right.\left|\mathbf{r}_{1},...,\mathbf{r}_{N};\mathbf{v}_{1},...,\mathbf{v}_{N}\right\rangle =\prod_{i=1}^{N}\delta(\mathbf{r}_{i}-\mathbf{r}'_{i})\delta(\mathbf{v}_{i}-\mathbf{v}'_{i}).
\end{equation}
The classical states are then defined to be finite norm vectors given
by a linear combination of the form

\begin{equation}
\left|\psi\right\rangle =\int_{R^{6N}}\psi\left|\mathbf{r}_{1},...,\mathbf{r}_{N};\mathbf{v}_{1},...,\mathbf{v}_{N}\right\rangle \:d\mathbf{r}_{1}d\mathbf{v}_{1}\ldots d\mathbf{r}_{N}d\mathbf{v}_{N},\label{psi}
\end{equation}
where $\psi$ is an square integrable function in $\mathbb{R}^{6N}$.
The probability of finding the first particle around $\mathbf{r}_{1}$
with velocity $\mathbf{v}_{1}$, and so on for the other particles,
is given by the Born rule

\begin{equation}
P(\mathbf{r}_{1},...,\mathbf{r}_{N};\mathbf{v}_{1},...,\mathbf{v}_{N})=\left|\psi\right|^{2}.
\end{equation}

The kets $\left|\mathbf{r}_{1},...,\mathbf{r}_{N};\mathbf{v}_{1},...,\mathbf{v}_{N}\right\rangle $
are simultaneous eigenvectors of $3N$ position and $3N$ velocity
operators 
\begin{align*}
\widehat{X}_{i}\left|\mathbf{r}_{1},...,\mathbf{r}_{N};\mathbf{v}_{1},...,\mathbf{v}_{N}\right\rangle  & =X_{i}\left|\mathbf{r}_{1},...,\mathbf{r}_{N};\mathbf{v}_{1},...,\mathbf{v}_{N}\right\rangle ,\\
\widehat{V}_{j}\left|\mathbf{r}_{1},...,\mathbf{r}_{N};\mathbf{v}_{1},...,\mathbf{v}_{N}\right\rangle  & =V_{j}\left|\mathbf{r}_{1},...,\mathbf{r}_{N};\mathbf{v}_{1},...,\mathbf{v}_{N}\right\rangle .
\end{align*}
The position and velocity operators form a complete set of commuting
operators. We now postulate the existence of the conjugate set operators
$\widehat{\lambda}$ and $\widehat{\pi}$ by the relations\footnote{The set of of operators $(\widehat{X},\widehat{V},\widehat{\lambda},\widehat{\pi})$
is irreducible in the Hilbert space we are considering. This mean
that our classical particles do not have any internal degree of freedom.
This is a choice by us, nothing force us to do so, we do it for simplicity.}

\begin{eqnarray}
\left[\widehat{X}_{i},\widehat{X}_{j}\right] & = & \left[\widehat{X}_{i},\widehat{V}_{j}\right]=\left[\widehat{V}_{i},\widehat{V}_{j}\right]=0,\nonumber \\
\left[\widehat{X}_{i},\widehat{\lambda}_{j}\right] & = & \left[\widehat{V}_{i},\widehat{\pi}_{j}\right]=0,\nonumber \\
\left[\widehat{X}_{i},\widehat{\lambda}_{j}\right] & = & \left[\widehat{V}_{i},\widehat{\pi}_{j}\right]=i\delta_{ij}.\label{comm}
\end{eqnarray}
The dynamics of the state vectors is encoded in a time evolution operator,
the so-called Liouvillian operator

\begin{equation}
\widehat{L}=\sum_{i=1}^{3N}\widehat{V}_{i}\widehat{\lambda_{i}}+\sum_{i=1}^{3N}\frac{1}{m_{i}}\left(\widehat{F}_{i}\widehat{\pi}_{i}\right)_{+},\label{L0}
\end{equation}
where $\left(\hat{A}\hat{B}\right)_{+}=\frac{1}{2}\left(\hat{A}\hat{B}+\hat{B}\hat{A}\right)$,
and $m_{1}=m_{2}=m_{3}$ is the mass of the first particle, and a
similar convention is used for all other particles. In (\ref{L0}),
the force operators can be function on the position and the velocity.
The form of the Liouvillian can be deduced from symmetry principles
and some other quite reasonable assumptions \cite{free1,free2}.

In the Schr\"{o}dinger picture, the evolution of the state vectors is
given by the Schr\"{o}dinger-like equation

\begin{equation}
\frac{d}{dt}\left|\psi(t)\right\rangle =-i\widehat{L}\left|\psi(t)\right\rangle .\label{Schr\"{o}dinger}
\end{equation}
Just as in quantum mechanics, there is the possibility of using a
Heisenberg picture, where the time dependence is passed on to the
operator via the Heisenberg equations. In this case, the Heisenberg
equations read

\begin{align}
\frac{d}{dt}\widehat{X}_{i}(t) & =i\left[\widehat{L},\widehat{X}_{i}\right]=\widehat{V}_{i}(t),\nonumber \\
\frac{d}{dt}\widehat{V}_{i}(t) & =i\left[\widehat{L},\widehat{V}_{i}\right]=\frac{1}{m_{i}}\widehat{F}_{i},\label{VF}
\end{align}
with analogous equations for the operators $\widehat{\lambda}$ and
$\widehat{\pi}$.

Examples where the the Schr\"{o}dinger-like equation (\ref{Schr\"{o}dinger})
leads to the expected dynamics of a classical system can be found
in \cite{free1}.

\subsection{Lagrangian and canonical momentum operators.}

Let us split the forces as 
\begin{equation}
\widehat{F}_{j}=\widehat{F}_{j}^{(C)}+\widehat{F_{j}}^{(NC)},
\end{equation}
where, by definition, the components of $\widehat{F}_{j}^{(C)}$ can
be computed from a generalized potential $\hat{U}=\hat{U}(\hat{X},\hat{V})$
according to\footnote{This kind of forces are called monogenic by Lanczos \cite{lanczos}.}

\begin{equation}
\widehat{F}_{j}^{(C)}=-i\left[\widehat{\lambda}_{j},\hat{U}\right]-\left[\widehat{L},\left[\widehat{\pi}_{j},\hat{U}\right]\right],\label{fpotential}
\end{equation}
and then let us define the Lagrange operator by

\begin{equation}
\hat{\mathscr{L}}=\sum_{i=1}^{3N}\frac{m_{i}}{2}\hat{V}_{i}^{2}-\hat{U}.\label{lagrangeoperator}
\end{equation}
For forces that are independent on the acceleration, the generalized
potential is restricted to be of the form \cite{free1}

\begin{equation}
\hat{U}=\hat{\phi}-\sum_{i=1}^{3N}\widehat{V}_{i}\hat{A}_{i},
\end{equation}
where $\widehat{\phi}$ and $\widehat{A}_{i}$ are only on the position.

We can then rewrite Eq (\ref{VF}) as

\begin{equation}
\Phi_{j}[\hat{\mathscr{L}}]=\widehat{F}_{j}^{(NC)},\label{phiL}
\end{equation}
where $\Phi_{j}$ is the following family of superoperators

\begin{equation}
\Phi_{j}=-\left[\hat{L},\left[\widehat{\pi}_{j},\right]\right]-i\left[\hat{\lambda}_{j},\right].
\end{equation}
We will call $\Phi_{j}$ as the Euler-Lagrange superoperators.

It is natural to look for a definition of a canonical momentum operator
once we have the Lagrangian (\ref{lagrangeoperator}). We define the
components of the canonical momentum by

\begin{equation}
\widehat{P}_{i}=i\left[\widehat{\pi}_{i},\widehat{\mathcal{L}}\right]=m_{i}\widehat{V}_{i}+\widehat{A}_{i}.\label{defP}
\end{equation}
It is straightforward to check the following commutation relations
for $\widehat{\mathbf{P}}$

\begin{eqnarray}
\left[\widehat{X}_{i},\widehat{P}_{j}\right] & = & 0,\\
\left[\widehat{P}_{i},\widehat{\pi}_{j}\right] & = & im_{i}\delta_{ij},\\
\left[\widehat{P}_{i},\widehat{\lambda}_{j}\right] & = & i\frac{\partial\widehat{A}_{i}}{\partial\widehat{X}_{j}}.
\end{eqnarray}

The introduction of a canonical momentum allows the passage from a
position-velocity description of the dynamics to a position-momentum
one. Let us consider the transformation equations

\begin{eqnarray}
\widehat{\pi}'_{i} & = & \frac{1}{m_{i}}\widehat{\pi}_{i},\label{changelambda1}\\
\widehat{\lambda}'_{j} & = & \widehat{\lambda}_{j}-\sum_{particle}\frac{\partial\widehat{A}_{i}}{\partial\widehat{X}_{j}}\widehat{\pi}'_{i}.\label{changelambda2}
\end{eqnarray}
The commutation relations between the operators $\widehat{X}$, $\widehat{P}$,
$\widehat{\lambda}'$ and $\widehat{\pi}'$ are

\begin{eqnarray}
\left[\widehat{X}_{i},\widehat{X}_{j}\right] & = & \left[\widehat{X}_{i},\widehat{P}_{j}\right]=\left[\widehat{P}_{i},\widehat{P}_{j}\right]=0,\nonumber \\
\left[\widehat{X}_{i},\widehat{\pi}'_{j}\right] & = & \left[\widehat{P}_{i},\widehat{\lambda}'_{j}\right]=0,\nonumber \\
\left[\widehat{X}_{i},\widehat{\lambda}'_{j}\right] & = & \left[\widehat{P}_{i},\widehat{\pi}'_{j}\right]=i\delta_{ij}.\label{X}
\end{eqnarray}

Hence, because of the commutation relations (\ref{X}), we can conclude
that the Eqs. (\ref{defP}), (\ref{changelambda1}), and (\ref{changelambda2})
form a ``quantum'' canonical transformation as understood in \cite{QCT}.
This transformation is a composition of a scale transformation and
a unitary transformation\cite{free1}. It can be checked that the
unitary operator

\begin{equation}
\widehat{C}=\exp\left(\sum_{i=1}^{3N}\frac{i}{m_{i}}\widehat{A}_{i}\widehat{\pi}'_{i}\right),
\end{equation}
gives the following 

\begin{align}
\widehat{C}\widehat{X}_{i}\widehat{C}^{-1} & =\widehat{X}_{i},\\
\widehat{C}(m_{i}\widehat{V}_{i})\widehat{C}^{-1} & =\widehat{P}_{i},\\
\widehat{C}\widehat{\lambda}{}_{j}\widehat{C}^{-1} & =\widehat{\lambda}'_{j},\\
\widehat{C}\widehat{\pi}{}_{j}\widehat{C}^{-1} & =\widehat{\pi}'_{j}.
\end{align}

In terms of the momentum (\ref{defP}) and the operators (\ref{changelambda1})
and (\ref{changelambda2}), the Liouvillian read

\begin{eqnarray}
\widehat{L} & = & \sum_{i=1}^{3N}\frac{1}{m_{i}}\left(\widehat{P}_{i}-\widehat{A}_{i}\right)\widehat{\lambda}'_{i}+\sum_{i=1}^{3N}\left(\widehat{F}_{i}\widehat{\pi}'_{i}\right)_{+}\nonumber \\
 &  & +\frac{1}{m}\frac{\partial\widehat{A}_{j}}{\partial\widehat{X}_{i}}\sum_{i,j=1}^{3N}\left\{ \left(\widehat{P}_{i}-\widehat{A}_{i}\right)\widehat{\pi}'_{j}\right\} _{+}.\label{LP}
\end{eqnarray}
The procedure to obtain equation (\ref{LP}) explains from first principles
the origin of the minimal coupling in the KvN theory given in \cite{minimal coupling}.
We can check that the Liouvillian (\ref{LP}) is consistent with (\ref{VF})
as follows

\begin{eqnarray}
i\left[\widehat{L},\widehat{V}_{i}\right] & = & i\left[\widehat{L}',\frac{1}{m}\left(\widehat{P}_{i}-\widehat{A}_{i}\right)\right]=\frac{1}{m}\widehat{F}_{i},\label{LVF-1-1}\\
i\left[\widehat{L},\widehat{X}_{i}\right] & = & i\left[\widehat{L}',\widehat{X}_{i}\right]=\frac{1}{m}\left(\widehat{P}_{i}-\widehat{A}_{i}\right)=\widehat{V}_{i}.\label{RV2-2}
\end{eqnarray}
Contrary to Ref \cite{free1}, we make the definition
\begin{equation}
\left|\mathbf{r}_{1},...,\mathbf{r}_{N};\mathbf{p}_{1},...,\mathbf{p}_{N}\right\rangle =\left|\mathbf{r}_{1},...,\mathbf{r}_{N};\mathbf{v}_{1},...,\mathbf{v}_{N}\right\rangle .
\end{equation}

Finally, it can be shown that a wave mechanics version of the operational
theory developed in this section leads to the KvN theory, and, ultimately,
to Hamiltonian mechanics. 

\section{Time-independent Holonomic constraints }

Consider a classical system composed by $N$ point particles that
obey the $l$ independent holonomic constraints 

\begin{equation}
f_{i}(x_{1},...,x_{3N})=C_{i},\quad(i=1,\ldots,l).\label{holonomic0}
\end{equation}
where the $C_{i}$ are constants. The Eq (\ref{holonomic0}) can be
equivalently written as

\begin{equation}
\sum_{j}\frac{\partial f_{i}}{\partial x_{j}}v_{j}=0\quad(i=1,\ldots,l).\label{Holonomic0V}
\end{equation}
Equations (\ref{holonomic0}) define a $3N-l$ dimensional submanifold
embedded on $\mathbb{R}^{3N}$, the configuration space $Q$. On the
other hand, the equations (\ref{holonomic0}) and (\ref{Holonomic0V})
together define a $6N-2l$ dimensional submanifold embedded on $\mathbb{R}^{6N}$,
the tangent bundle of configuration space $TQ$.

What is the meaning of a holonomic constraint in the operational version
of mechanics? We will interpret the constraints as a restriction on
the possible states $\left|\psi\right\rangle $ that our mechanical
system can be, such that the allowed states belong to a subset of
the unconstrained Hilbert space $\mathcal{H}_{TQ}\subseteq\mathcal{H}$.
We postulate that the allowed states $\left|\psi_{TQ}\right\rangle $
obey an operational versions of (\ref{holonomic0}) and (\ref{Holonomic0V})
given by

\begin{align}
\left\langle \psi_{TQ}\right|f_{i}(\widehat{X}_{1},..,\widehat{X}_{3N})\left|\psi_{TQ}\right\rangle  & =C_{i},\label{holonomic}\\
\left\langle \psi_{TQ}\right|\sum_{j}\frac{\partial f_{i}}{\partial\widehat{X}_{j}}\widehat{V}_{j}\left|\psi_{TQ}\right\rangle  & =0\quad(i=1,\ldots,l).\label{holonomicv}
\end{align}
The set $\mathcal{H}_{TQ}\subseteq\mathcal{H}$ containing all the
vectors that obey (\ref{holonomic}) is a linear subsapce of $\mathcal{H}$
since $\mathcal{H}_{TQ}$ is clearly non empty ($0\in\mathcal{H}_{TQ}$),
and we have that $\left|\psi_{TQ1}\right\rangle +\lambda\left|\psi_{TQ2}\right\rangle \in\mathcal{H}_{TQ}$
for any $\left|\psi_{TQ1}\right\rangle ,\left|\psi_{TQ2}\right\rangle \in\mathcal{H}_{c}$.

Let us now investigate the form of the vectors belonging to $\mathcal{H}_{TQ}$.
By our initial definition (\ref{psi}), we can write any vector $\left|\psi_{TQ1}\right\rangle \in\mathcal{H}_{TQ}\subseteq\mathcal{H}$
as

\begin{equation}
\left|\psi_{TQ}\right\rangle =\int_{R^{6N}}\psi\left|\mathbf{r}_{1},...,\mathbf{r}_{N};\mathbf{v}_{1},...,\mathbf{v}_{N}\right\rangle \:d\mathbf{r}_{1}d\mathbf{v}_{1}\ldots d\mathbf{r}_{N}d\mathbf{v}_{N},\label{psic}
\end{equation}
for some function $\psi=\psi(\mathbf{r}_{1},...,\mathbf{r}_{N};\mathbf{v}_{1},...,\mathbf{v}_{N}).$
Replacing (\ref{psic}) into (\ref{holonomic}) and (\ref{holonomicv})
we get the set of equations

\begin{align}
0 & =\int_{R^{6N}}\left|\psi\right|^{2}\:f_{i}\,d\mathbf{r}_{1}d\mathbf{v}_{1}\ldots d\mathbf{r}_{N}d\mathbf{v}_{N},\nonumber \\
0 & =\int_{R^{6N}}\left|\psi\right|^{2}\:\left\{ \sum_{j}\frac{\partial f_{i}}{\partial x_{j}}v_{j}\right\} \,d\mathbf{r}_{1}d\mathbf{v}_{1}\ldots d\mathbf{r}_{N}d\mathbf{v}_{N}\quad(i=1,\ldots,n).\label{constraintintegral}
\end{align}
Since $\left|\psi\right|^{2}$ is a non negative everywhere, $\psi$
must vanish for the values of the positions and velocities that do
not obey (\ref{holonomic0}) and (\ref{Holonomic0V}) for Eq. (\ref{constraintintegral})
to be true. This is, the support of $\psi$ is $TQ$, and we can write
\[
\psi=\left\{ \begin{array}{c}
\psi_{TQ}\:\:\mathrm{for}\:\mathrm{points}\:\mathrm{in}\:TQ,\\
0\quad\mathrm{outside}\:TQ.\quad\;
\end{array}\right.
\]
Equivalently, we can write our vectors $\left|\psi_{TQ}\right\rangle $
by a restriction on the domain of integration
\begin{equation}
\left|\psi_{TQ}\right\rangle =\int_{TQ}\psi_{TQ}\left|\mathbf{r}_{1},...,\mathbf{r}_{N};\mathbf{v}_{1},...,\mathbf{v}_{N}\right\rangle \:d\mathbf{r}_{1}d\mathbf{v}_{1}\ldots d\mathbf{r}_{N}d\mathbf{v}_{N}.
\end{equation}

\section{Generalized coordinates}

It is always possible to find a set of $n=3N-l$ independent generalized
coordinates $(q_{1},...,q_{n})$ to fully describe a mechanical system
with $l$ holonomic constraints. It is not difficult to pass from
wavefunctions on cartesian coordinates to wave functions on generalized
coordinates: $\psi(\mathbf{r}_{1},...,\mathbf{r}_{N};\mathbf{v}_{1},...,\mathbf{v}_{N})\rightarrow\psi(q_{1},...,q_{n};\dot{q}_{1},...,\dot{q}_{n})$.
However, we will work with operators since we want an operational
version of mechanics.

Let us review the (time-independent) transformation of operators to
curvilinear coordinates as given in \cite{dewitt}, and then we shall
mention how to apply the procedure to coordinates in $TQ$. Since
the original position and velocity operators commute with each other,
the point transformation to curvilinear coordinates can be defined
in an unambiguous manner\footnote{Notice that here we are explicitly dealing with time-independent transformation
of coordinates.}

\begin{align}
\hat{q}_{j} & =\hat{q}_{j}(\widehat{X}_{1},...,\widehat{X}_{3N}),\label{Gq}\\
\hat{v}_{j} & =\sum_{i=1}^{3N}\widehat{V}_{i}\frac{\partial\widehat{q}_{j}}{\partial\widehat{X}_{i}}\quad(j=1,\ldots,3N).\label{Gv}
\end{align}
The above procedure works as long as the spectrum of the $\hat{q}$
runs from $-\infty$ to $+\infty$. Angular coordinates are more delicate,
and we will deal with them later. From Eqs (\ref{Gq}) and (\ref{Gv})
and the commutator relations (\ref{comm}), it follows that 

\begin{equation}
\left[\widehat{q}_{j},\widehat{v}_{i}\right]=\left[\widehat{v}_{j},\widehat{v}_{i}\right]=\left[\widehat{q}_{j},\widehat{v}_{i}\right]=0.
\end{equation}

We now insist that the transformation rule for the auxiliary operators
$\widehat{\lambda}$ and $\widehat{\pi}$ is such that the point transformation
is a ``quantum'' canonical transformation. In analogy with the definitions
given in \cite{dewitt}, we give the following rule

\begin{align}
\hat{\lambda}_{i}^{(q)} & =\sum_{i=1}^{3N}\left(\frac{\partial\hat{X}_{j}}{\partial\hat{q}_{i}}\hat{\lambda}_{j}\right)_{+}+\sum_{i=1}^{3N}\left(\frac{\partial\hat{V}_{j}}{\partial\hat{q}_{i}}\hat{\pi}_{j}\right)_{+},\label{point1}\\
\hat{\pi}_{i}^{(q)} & =\sum_{i=1}^{3N}\left(\frac{\partial\hat{V}_{j}}{\partial\hat{v}_{i}}\hat{\pi}_{j}\right)_{+}.\label{point2}
\end{align}
Notice that a term $\sum_{i=1}^{3N}\left(\frac{\partial\hat{X}_{i}}{\partial\hat{v}_{j}}\hat{\lambda}_{j}\right)_{+}$
is missing from Eq (\ref{point2}) because the inverse transformations
$\widehat{X}_{i}=\widehat{X}_{i}(\hat{q}_{1},...,\hat{q}_{j})$ are
independent of the $\hat{v}$. We can then verify that the following
commutator relations are obeyed

\begin{align}
\left[\widehat{q}_{i},\hat{\pi}_{j}^{(q)}\right] & =\left[\widehat{v}_{i},\hat{\lambda}_{j}^{(q)}\right]=\left[\hat{\lambda}_{i}^{(q)},\hat{\pi}_{j}^{(q)}\right]=0,\nonumber \\
\left[\widehat{q}_{i},\hat{\lambda}_{j}^{(q)}\right] & =\left[\widehat{v}_{i},\hat{\pi}_{j}^{(q)}\right]=i.\label{comm2}
\end{align}

The set $\left\{ \widehat{q},\widehat{v}\right\} $ is a complete
set of commuting operators. Hence, there exist a basis of common eigenkets
that can be labeled via their eigenvalues, this is, there exist a
basis of kets $\left|\widehat{q}_{1},\ldots,\widehat{q}_{3N};\widehat{v}_{1},\ldots,\widehat{v}_{3N}\right\rangle $
such that

\begin{align}
\widehat{q}_{j}\left|\widehat{q}_{1},\ldots,\widehat{q}_{3N};\widehat{v}_{1},\ldots,\widehat{v}_{3N}\right\rangle  & =q_{j}\left|\widehat{q}_{1},\ldots,\widehat{q}_{3N};\widehat{v}_{1},\ldots,\widehat{v}_{3N}\right\rangle ,\\
\widehat{v}_{j}\left|\widehat{q}_{1},\ldots,\widehat{q}_{3N};\widehat{v}_{1},\ldots,\widehat{v}_{3N}\right\rangle  & =v_{j}\left|\widehat{q}_{1},\ldots,\widehat{q}_{3N};\widehat{v}_{1},\ldots,\widehat{v}_{3N}\right\rangle .
\end{align}
It so happen that the ket $\left|\widehat{q}_{1},\ldots,\widehat{q}_{3N};\widehat{v}_{1},\ldots,\widehat{v}_{3N}\right\rangle $
is proportional to $\left|\mathbf{r}_{1},...,\mathbf{r}_{N};\mathbf{v}_{1},...,\mathbf{v}_{N}\right\rangle $
as long as their label correspond to the same point in $T\mathbb{R}^{3N}$.
We must have the relation \cite{dewitt}

\[
\left|\widehat{q}_{1},\ldots,\widehat{q}_{3N};\widehat{v}_{1},\ldots,\widehat{v}_{3N}\right\rangle =\left|J\right|^{-1/2}\left|\mathbf{r}_{1},...,\mathbf{r}_{N};\mathbf{v}_{1},...,\mathbf{v}_{N}\right\rangle ,
\]
where $\left|J\right|$ is the Jacobian determinant of the transformation.
The new basis kets obey the orthonormality condition

\[
\left\langle \widehat{q}'_{1},\ldots,\widehat{q}'_{3N};\widehat{v}'_{1},\ldots,\widehat{v}'_{3N}\right.\left|\widehat{q}_{1},\ldots,\widehat{q}_{3N};\widehat{v}_{1},\ldots,\widehat{v}_{3N}\right\rangle =\left|J\right|^{-1}\prod_{i=1}^{3N}\delta(q_{i}-q'_{i})\delta(v_{i}-v'_{i}).
\]
Finally, a general state can be written using the new coordinates
as

\begin{equation}
\left|\psi\right\rangle =\int_{R^{6N}}\psi^{(q)}\left|\widehat{q}_{1},\ldots,\widehat{q}_{3N};\widehat{v}_{1},\ldots,\widehat{v}_{3N}\right\rangle \:\left|J\right|dq_{1}dv_{1}\ldots dq_{3N}dv_{3N},
\end{equation}

\subsection{Reduction of variables}

In analogy to analytical mechanics, we want to be able to describe
a holonomic system of $n$ degrees of freedom using just $n$ position
operators and $n$ velocity operators. Consider the following point
transformation

\begin{align}
\widehat{q}_{i} & =\widehat{q}_{i}(\widehat{X}_{j},...,\widehat{X}_{3N})\quad(i=1,\ldots,n),\\
\widehat{q}_{j} & =f_{j}(\widehat{X}_{1},..,\widehat{X}_{3N})\quad(j=n+1,\ldots,3N).\label{q=00003Df}
\end{align}

The expectation value of the coordinates $\widehat{q}_{j}\,(j=n+1,\ldots,3N)$
is constant for all the states that obey the constraints (\ref{holonomic})

\begin{equation}
\left\langle \psi_{TQ}\right|\widehat{q}_{j}\left|\psi_{TQ}\right\rangle =\left\langle \psi_{TQ}\right|f_{j}(\widehat{X}_{1},..,\widehat{X}_{3N})\left|\psi_{TQ}\right\rangle =C_{j}\quad(j=n+1,\ldots,3N).
\end{equation}
From (\ref{holonomicv}) and (\ref{Gv}), it follows that there are
generalized velocity operators with vanishing expectation value

\begin{equation}
\left\langle \psi_{TQ}\right|\widehat{v}_{j}^{\perp}\left|\psi_{TQ}\right\rangle =\left\langle \psi_{TQ}\right|\sum_{j}\frac{\partial f_{i}}{\partial\widehat{X}_{j}}\widehat{V}_{j}\left|\psi_{TQ}\right\rangle =0\quad(j=n+1,\ldots,3N).
\end{equation}
We call these $\widehat{v}^{\perp}$ as the perpendicular velocity
operators. 

From the above discussion we conclude that the vectors$\left|\psi_{TQ}\right\rangle $
are linear supperposition of kets of the form $\left|\widehat{q}_{1},\ldots,\widehat{q}_{n},C_{n+1},...,C_{N};\widehat{v}_{1},\ldots,\widehat{v}_{n},0...,0\right\rangle $.

Alternatively, we can consider an intrinsic point of view where $TQ$
is not understood as a submanifold of $\mathbb{R}^{6N}$ but as a
independent space of its own. Locally, to a point $(q_{1},...,q_{n},v_{1},\ldots,v_{n})$
in $TQ$, we can assign a ket of the form $\left|\widehat{q}_{1},\ldots,\widehat{q}_{n};\widehat{v}_{1},\ldots,\widehat{v}_{n}\right\rangle $
such that

\begin{align}
\widehat{q}_{j}\left|\widehat{q}_{1},\ldots,\widehat{q}_{n};\widehat{v}_{1},\ldots,\widehat{v}_{n}\right\rangle  & =q_{j}\left|\widehat{q}_{1},\ldots,\widehat{q}_{n};\widehat{v}_{1},\ldots,\widehat{v}_{n}\right\rangle ,\nonumber \\
\widehat{v}_{j}\left|\widehat{q}_{1},\ldots,\widehat{q}_{n};\widehat{v}_{1},\ldots,\widehat{v}_{n}\right\rangle  & =v_{j}\left|\widehat{q}_{1},\ldots,\widehat{q}_{n};\widehat{v}_{1},\ldots,\widehat{v}_{n}\right\rangle \quad(j=1,\ldots,n).
\end{align}
The Hilbert space spanned by the kets $\left|\widehat{q}_{1},\ldots,\widehat{q}_{n};\widehat{v}_{1},\ldots,\widehat{v}_{n}\right\rangle $
is different from the one we were considering before since it is build
from the tensor product of fewer one-particle spaces. However, there
is a one-to-one correspondence between the rays \{$\left|\widehat{q}_{1},\ldots,\widehat{q}_{n};\widehat{v}_{1},\ldots,\widehat{v}_{n}\right\rangle $\}
and \{$\left|\widehat{q}_{1},\ldots,\widehat{q}_{n},C_{n+1},...,C_{N};\widehat{v}_{1},\ldots,\widehat{v}_{n},0...,0\right\rangle $\}.
Abusing a bit the terminology, we can consider the elements of$\mathcal{H}_{TQ}$
as being given by

\begin{align}
\left|\psi_{TQ}\right\rangle  & =\int_{TQ}\psi(q,v)\left|\widehat{q}_{1},\ldots,\widehat{q}_{n};\widehat{v}_{1},\ldots,\widehat{v}_{n}\right\rangle \left|J\right|\:dq_{1}dv_{1}\ldots,dq_{n}dv_{n}.\label{psitq}
\end{align}
We arrive at the desired conclusion that the state of a system of
$n$ degree of freedom can be described using just $n$ position and
$n$ velocity variables (operators). We will see in section 5 that
for the dynamics is also enough to define the states by (\ref{psitq})
(as long as we are no interested in the constraining forces).

\subsection{Angle coordinates }

That angle operators in quantum mechanics are problematic has been
known for a long time\cite{angle operators}. The same applies to
our operational version of classical mechanics and for the same reasons,
most importantly, the cyclic nature of their expected spectra. We
need to modify the procedures of the preceding sections to accommodate
angle variables. The easiest way to deal with angle variables is to
use the sine and cosine function, such that instead of defining the
angle operator itself as $\hat{\theta}_{i}=\hat{\theta}_{i}(\widehat{X}_{1},...,\widehat{X}_{3N})$
we define instead the following Hermitian operators by their action
on the base kets \cite{angle operators 2}

\begin{align}
\sin\hat{\theta}_{i}\left|...\theta_{i}...\right\rangle  & =\sin\theta_{i}\left|...\theta_{i}...\right\rangle ,\\
\cos\hat{\theta}_{i}\left|...\theta_{i}...\right\rangle  & =\cos\theta_{i}\left|...\theta_{i}...\right\rangle .
\end{align}
To reiterate, the angle operators are well defined only when they
are arguments of the trigonometric functions. We shall never work
with angle operators by themselves.

We also need to define the conjugate operators to complete the transformation
of coordinates. By definition, they obey the commutation relation
\begin{align}
[\sin\hat{\theta}_{i},\hat{\lambda}_{i}] & =i\cos\hat{\theta}_{i},\\{}
[\cos\hat{\theta}_{i},\hat{\lambda}_{i}] & =-i\sin\hat{\theta}_{i}.
\end{align}
It is worth pointing out that there is no problem with the associated
velocities. To each angle coordinate we can associate a velocity operator
$\hat{v}_{\theta i}$ with spectrum between $-\infty$ and $\infty$,
and its corresponding conjugate operator $\hat{\pi}_{i}$ is given
by (\ref{point2}).

In the appendix we work out in detail the change of variables from
Cartesian to spherical coordinates, and we use the result in there
to find the equation of motion of the spherical pendulum in section
5.1.

\section{Operational Dynamics in $TQ$}

In this section, we will express the Liouvillian in generalized coordinates,
and in doing so we will find that the Heisenberg equations for the
operators lead to the Lagrange equation of motion and the equation
for the constraint forces.We will designate by $\widehat{F}_{j}$
the sum of all the impressed (non-constraint forces) and $\widehat{R}_{j}$
stand for the sum of all the constraint forces acting on the $j-$particle.
As is usually the case in mechanics, we assume that the constraining
forces are not known a priori, their role is to maintain valid the
constraints. 

Consider the full Liouvillian of a mechanical system in Cartesian
coordinates

\begin{equation}
\widehat{L}=\sum_{i=1}^{3N}\widehat{V}_{i}\widehat{\lambda_{i}}+\sum_{i=1}^{3N}\frac{1}{m_{i}}\left(\widehat{F}_{i}\widehat{\pi}_{i}\right)_{+}+\sum_{i=1}^{3N}\frac{1}{m_{i}}\left(\widehat{R}_{i}\widehat{\pi}_{i}\right)_{+},
\end{equation}
Changing to generalized coordinates and velocities the Liouvillian
becomes
\begin{align}
\widehat{L} & =\sum_{i,j,k=1}^{3N}(\widehat{v}_{k}\frac{\partial\widehat{X}_{i}}{\partial\widehat{q}_{k}})\left(\left(\frac{\partial\hat{q}_{j}}{\partial\hat{X}_{i}}\hat{\lambda}_{j}^{(q)}\right)_{+}+\left(\frac{\partial\hat{v}_{j}}{\partial\hat{X}_{i}}\hat{\pi}_{j}^{(q)}\right)_{+}\right)+\sum_{i,j=1}^{3N}\frac{1}{2m_{i}}\left(\widehat{F}_{i}\left(\frac{\partial\hat{v}_{j}}{\partial\hat{V}_{i}}\hat{\pi}_{j}^{(q)}\right)_{+}\right)_{+}\nonumber \\
 & +\sum_{i=1}^{3N}\sum_{j=n+1}^{3N}\frac{1}{2m_{i}}\left(\widehat{R}_{i}\left(\frac{\partial\hat{v}_{i}}{\partial\hat{V}_{j}}\hat{\pi}_{j}^{(q)}\right)_{+}\right)_{+}.\label{Lq}
\end{align}
Since the mean value of the $\widehat{q}_{j}$ and $\widehat{v}_{j}$
for $j=n+1,\ldots,3N$ has to remain constant due to the constraints,
we must have
\begin{align}
\frac{d}{dt}\widehat{q}_{j}(t) & =i\left[\widehat{L},\widehat{q}_{j}\right]=\sum_{i,k=1}^{3N}(\widehat{v}_{k}\frac{\partial\widehat{X}_{i}}{\partial\widehat{q}_{k}})\frac{\partial\hat{q}_{j}}{\partial\hat{X}_{i}}=\widehat{v}_{j}(t)=0,\label{qv}\\
\frac{d}{dt}\widehat{v}_{j}(t) & =i\left[\widehat{L},\widehat{v}_{j}\right]=0,\label{dv}
\end{align}
where we used the inverse function theorem $\sum_{i=1}^{3N}\frac{\partial\widehat{X}_{i}}{\partial\widehat{q}_{k}}\frac{\partial\hat{q}_{i}}{\partial\hat{X}_{i}}=\delta_{kj}$,
and the equation is valid when evaluated in vectors belonging to $\mathcal{H}_{TQ}$.
Equation (\ref{qv}) gives an identity that does not give any new
information. On the other hand, Eq (\ref{dv}) gives the form of the
constraining forces
\begin{equation}
0=\sum_{k=1}^{n}\sum_{i,j=1}^{3N}\frac{\partial\widehat{X}_{i}}{\partial\widehat{q}_{k}}\frac{\partial\hat{v}_{j}}{\partial\hat{X}_{i}}\widehat{v}_{k}+\sum_{i=1}^{3N}\frac{1}{m_{i}}\widehat{F}_{i}\frac{\partial\hat{v}_{j}}{\partial\hat{V}_{i}}+\sum_{i=1}^{3N}\frac{1}{m_{i}}\widehat{R}_{i}\frac{\partial\hat{v}_{j}}{\partial\hat{V}_{i}}.\label{Rforces}
\end{equation}
The above system give $3N-l$ equations for the $3N-l$ unknown constraining
forces. We will return to analyze Eq (\ref{Rforces}) later, let us
now focus on the equation of motion for the remaining variables.

In view of Eq (\ref{qv}) and (\ref{dv}) we can conclude that, when
evaluated in $\mathcal{H}_{TQ}$, $\widehat{L}$ is independent of
$\hat{\lambda}_{j}^{(q)}$ and $\hat{\pi}_{j}^{(q)}$ for $j=n+1,\ldots,3N$.
Hence, we can rewrite Eq (\ref{Lq}) as 

\begin{align}
\widehat{L} & =\sum_{k=1}^{n}\widehat{v}_{k}\hat{\lambda}_{k}^{(q)}+\sum_{i,j,k=1}^{3N}\widehat{v}_{k}(\frac{\partial\widehat{X}_{i}}{\partial\widehat{q}_{k}})\frac{\partial\hat{v}_{j}}{\partial\hat{X}_{i}}\hat{\pi}_{j}^{(q)}\nonumber \\
 & +\sum_{i,j=1}^{3N}\frac{1}{m_{i}}\widehat{F}_{i}\frac{\partial\hat{v}_{j}}{\partial\hat{V}_{i}}\hat{\pi}_{j}^{(q)}+\hat{f}.\label{L3}
\end{align}
where $\hat{f}$ is an operator independent of $\hat{\lambda}^{(q)}$
and $\hat{\pi}^{(q)}$ obtained from simplifying the anticommutators
in (\ref{Lq}). The operator $\hat{f}$ has no effect on the equation
of motion of the position and velocity operators, its main function
is to guarantee the Hermiticity of $\widehat{L}$.

The Heisenberg equation of motion obtained from (\ref{L3}) are

\begin{align}
\frac{d}{dt}\widehat{q}_{j}(t) & =i\left[\hat{L},\widehat{q}_{j}\right]=\widehat{v}_{j}(t),\label{qv-1}\\
\frac{d}{dt}\widehat{v}_{j}(t) & =i\left[\hat{L},\widehat{v}_{j}\right]=\sum_{i,k=1}^{3N}\frac{\partial\widehat{X}_{i}}{\partial\widehat{q}_{k}}\frac{\partial\hat{v}_{j}}{\partial\hat{X}_{i}}\widehat{v}_{k}+\sum_{,i=1}^{3N}\frac{1}{m_{i}}\widehat{F}_{i}\frac{\partial\hat{v}_{j}}{\partial\hat{V}_{i}}\;\left(j=1,...,n\right),\label{dv-1}
\end{align}
 Now, the system of $n$ equations (\ref{dv-1}) can be rewritten
as

\begin{equation}
\sum_{j=1}^{n}\sum_{,l=1}^{3N}m_{l}\frac{\partial\widehat{X}_{l}}{\partial\widehat{q}_{i}}\frac{\partial\widehat{X}_{l}}{\partial\widehat{q}_{j}}\frac{d}{dt}\widehat{v}_{j}+\sum_{j,k=1}^{n}\sum_{l=1}^{3N}m_{l}(\frac{\partial\widehat{X}_{l}}{\partial\widehat{q}_{i}}\frac{\partial^{2}\widehat{X}_{l}}{\partial\widehat{q}_{j}\partial\widehat{q}_{k}})\hat{v}_{j}\widehat{v}_{k}=\sum_{l=1}^{3N}\widehat{F}_{l}\frac{\partial\widehat{X}_{l}}{\partial\widehat{q}_{i}}\:\left(i=1,...,n\right),\label{Lequations}
\end{equation}
where the following identities have been used

\begin{align}
\frac{\partial\hat{V}_{j}}{\partial\hat{v}_{i}} & =\frac{\partial\hat{X}_{j}}{\partial\hat{q}_{i}},\label{I1}\\
\frac{\partial\hat{v}_{j}}{\partial\hat{X}_{i}} & =\sum_{k=1}^{n}\frac{\partial\hat{X}_{l}}{\partial\widehat{q}_{k}}\frac{\partial^{2}\widehat{q}_{j}}{\partial\hat{X}_{i}\partial\hat{X}_{l}}\widehat{v}_{k}\:(\widehat{v}_{k}=0\:for\:k=n+1,...,3N),\label{I2}\\
\frac{\partial^{2}\hat{X}_{l}}{\partial\widehat{q}_{j}\partial\widehat{q}_{k}} & =-\sum_{\alpha,\beta,\gamma=1}^{3N}\frac{\partial\hat{X}_{l}}{\partial\widehat{q}_{\alpha}}\frac{\partial\hat{X}_{\beta}}{\partial\widehat{q}_{k}}\frac{\partial\hat{X}_{\gamma}}{\partial\widehat{q}_{j}}\frac{\partial^{2}\widehat{q}_{\alpha}}{\partial\hat{X}_{\beta}\partial\hat{X}_{\gamma}}.\label{I3}
\end{align}
The equation (\ref{I1}) is a common identity in Lagrangian mechanics\cite[Eq (2-15)]{Schwinger},
Eq (\ref{I2}) comes from the definition of generalized velocities,
and Eq (\ref{I3}) comes from the properties of the Hessian matrix.

The systems of equations (\ref{Lequations}) is the same as the one
obtained from Lagrangian mechanics (compare with \cite[Eq (2-37)]{Schwinger}).

Let us now consider equation (\ref{Rforces}) in more details. Using
identities (\ref{I1}) and (\ref{I1}), we can solve (\ref{Rforces})
for the force of constraint as as 
\begin{equation}
\widehat{R}_{i}^{(gen)}=\sum_{j,k=1}^{n}\sum_{l=1}^{3N}m_{l}(\frac{\partial\widehat{X}_{l}}{\partial\widehat{q}_{i}}\frac{\partial^{2}\widehat{X}_{l}}{\partial\widehat{q}_{j}\partial\widehat{q}_{k}})\hat{v}_{j}\widehat{v}_{k}-\sum_{l=1}^{3N}\widehat{F}_{l}\frac{\partial\widehat{X}_{l}}{\partial\widehat{q}_{i}}\:\left(i=n+1,...,3N\right)\label{Rgen}
\end{equation}
where the components of the generalized constraint force $\widehat{R}_{i}^{(gen)}$
are related to the Cartesian components via 
\begin{equation}
\widehat{R}_{i}^{(gen)}=\sum_{l=1}^{3N}\widehat{R}_{l}\frac{\partial\widehat{X}_{l}}{\partial\widehat{q}_{i}}.\label{rgen}
\end{equation}

\subsection{Example: Spherical pendulum}

A spherical pendulum consist of a mass restricted to move on the surface
of a sphere of radius $r$ under the action of the gravity. The constraining
force in this case is the tension that is always in the radial direction.
The complete Liouvillian of the system in Cartesian coordinates is

\begin{equation}
\hat{L}=\widehat{\mathbf{V}}\cdot\widehat{\lambda}+\frac{1}{m}(\widehat{\mathbf{T}}\cdot\widehat{\pi})_{+}-g\widehat{\pi}_{z}.
\end{equation}

We are going to write the reduced Liouvillian taking into account
the constraints

\begin{align}
\left\langle \psi_{TQ}\right|\left(\widehat{x}^{2}+\widehat{y}^{2}+\widehat{z}^{2}\right)\left|\psi_{TQ}\right\rangle  & =r^{2},\nonumber \\
\left\langle \psi_{TQ}\right|\widehat{V}_{r}\left|\psi_{TQ}\right\rangle  & =0.\label{constraint 2}
\end{align}
and ignoring the constraining force $\widehat{\mathbf{T}}$. Using
the transformation equations (\ref{T1}), (\ref{T2}) (\ref{lambda})
and (\ref{pi}) we obtain

\begin{equation}
\hat{L}=\hat{V}_{\theta}\widehat{\lambda}_{\theta}+\hat{V}_{\phi}\widehat{\lambda}_{\phi}+\sin\widehat{\theta}\cos\widehat{\theta}\,\widehat{V}_{\phi}^{2}\,\widehat{\pi}_{\theta}-2\hat{V}_{\phi}\hat{V}_{\theta}\cot\widehat{\theta}\,\widehat{\pi}_{\phi}-g\frac{\sin\widehat{\theta}}{r}\widehat{\pi}_{\theta}+f(r,\widehat{\theta},\widehat{\phi}),
\end{equation}
where $f$ is independent on $\hat{\lambda}$ and $\hat{\pi}$, and,
in view of the constraints (\ref{constraint 2}), we have made the
substitutions $\hat{r}\rightarrow r$ and $\hat{V_{r}}=0.$ We reiterate
that this substitutions are only valid when the Liouvillian is restricted
to act on vectors that obey (\ref{constraint 2}). 

The Heisenberg equation for the velocities are
\begin{align}
\frac{d\hat{V}_{\theta}}{dt} & =i\left[\hat{L},\hat{V}_{\theta}\right]=\sin\widehat{\theta}\cos\widehat{\theta}\widehat{V}_{\phi}-\frac{g}{r}\sin\widehat{\theta},\nonumber \\
\frac{d\hat{V}_{\phi}}{dt} & =i\left[\hat{L},\hat{V}_{\phi}\right]=-2\hat{V}_{\phi}\hat{V}_{\theta}\cot\widehat{\theta}.\label{spherical1}
\end{align}
Now, the only constraining force and it is in the radial direction.
In view of (\ref{rgen}) and (\ref{T1}), we can write
\begin{equation}
\widehat{R}_{i}^{(gen)}=\hat{T_{x}}\frac{\partial\hat{x}}{\partial r}+\hat{T_{y}}\frac{\partial\hat{y}}{\partial r}+\hat{T_{z}}\frac{\partial\hat{z}}{\partial r}=\hat{T}.
\end{equation}
The generalized force is then the magnitude of the tension. Using
(\ref{Rgen}), the tension can then be calculated to be

\begin{equation}
\hat{T}=mr(\hat{V}_{\theta}^{2}+\hat{V}_{\phi}^{2}\sin\widehat{\theta})+mg\cos\widehat{\theta}.\label{spherical2}
\end{equation}

We see that the equation of motion (\ref{spherical1}) and the tension
(\ref{spherical2}) coincide with the ones obtained from Lagrangian
mechanics.

\subsection{Lagrangian and generalized momentum }

This section follows the same structure as section 2.1, and the objective
is the same, to pass from a position-velocity description of the movement
to a position-momentum one via a ``quantum'' canonical transformation. 

We start by rewriting Eq (\ref{Lequations}) as

\begin{equation}
\Phi_{i}[\hat{\mathcal{T}}]=\hat{Q}_{i},\label{QT1}
\end{equation}
where the kinetic energy operator is given by 
\begin{align}
\hat{\mathcal{T}} & =\frac{1}{2}\sum_{j,k=1}^{n}m_{jk}\widehat{v}_{j}\widehat{v}_{k},\nonumber \\
m_{jk} & =\sum_{,l=1}^{3N}m_{l}\frac{\partial\widehat{X}_{l}}{\partial\widehat{q}_{k}}\frac{\partial\widehat{X}_{l}}{\partial\widehat{q}_{j}},
\end{align}
the generalized forces are defined by

\begin{equation}
\hat{Q}_{i}=\sum_{l=1}^{3N}\widehat{F}_{l}\frac{\partial\widehat{X}_{l}}{\partial\widehat{q}_{i}},
\end{equation}
and the Euler-Lagrange superoperators are 
\begin{equation}
\Phi_{i}=-\left[\widehat{L},\left[\widehat{\pi}_{i}^{(q)},\right]\right]-i\left[\hat{\lambda}_{i}^{(q)},\right].
\end{equation}
For generalized forces that can be derived from a generalized potential
we have that

\begin{align}
\hat{Q}_{i}^{(C)} & =-i\left[\widehat{\lambda}_{j},\widehat{U}\right]-\left[\widehat{L},\left[\widehat{\pi}_{j},\widehat{U}\right]\right],\\
\hat{U} & =\hat{\phi}+\sum_{i=1}^{n}\hat{q}_{i}\hat{A}_{i}.
\end{align}
Equation (\ref{QT1}) can then be rewritten as 

\begin{align}
\Phi_{i}[\hat{\mathscr{L}}] & =\hat{Q}_{i}^{(NC)},\label{eulerlagrange}\\
\hat{\mathscr{L}} & =\hat{\mathcal{T}}-\hat{U}.\label{lagrangianoperator}
\end{align}
With the Lagrangian (\ref{lagrangianoperator}) already defined, we
can now define the momentum operators by

\begin{equation}
\hat{p}_{i}=i\left[\widehat{\pi}_{i}^{(q)},\hat{\mathscr{L}}\right]=\sum_{j=1}^{n}m_{ik}\widehat{v}_{k}+\hat{A}_{i}.\label{P}
\end{equation}
Just as in section 2.1, we demand that the passage to a momentum description
of the dynamics entails a ``quantum'' canonical transformation.
Consider the following transformation rules for the auxiliary operators 

\begin{align}
\widehat{\pi}'_{j} & =\sum_{l=1}^{n}m_{lj}^{-1}\widehat{\pi}_{l}^{(q)},\nonumber \\
\hat{\lambda}'_{j} & =\hat{\lambda}_{j}^{(q)}-\sum m_{l\alpha}^{-1}\left(\sum v_{\beta}\frac{\partial m_{\beta\alpha}}{\partial\widehat{q}_{j}}+\frac{\partial\hat{A}_{\alpha}}{\partial\widehat{q}_{j}}\right)\widehat{\pi}_{l}^{(q)}.\label{Prepresentation}
\end{align}
It can be checked that (\ref{P}) and (\ref{Prepresentation}) obey
the canonical commutation relations

\begin{align}
[\hat{q}_{i},\hat{\lambda}'_{j}] & =[\hat{p}_{i},\widehat{\pi}'_{j}]=i\delta_{ij},\nonumber \\{}
[\hat{q}_{i},\widehat{\pi}'_{j}] & =[\hat{p}_{i},\hat{\lambda}'_{j}]=[\widehat{\pi}'_{i},\hat{\lambda}'_{j}]=0.\label{pQtransformation}
\end{align}
Hence, the transformations (\ref{P}) and (\ref{Prepresentation})
give the desired ``quantum'' canonical transformation.

Let us investigate now the form of the Liouvillian as a function of
the momentum. For simplicity, we put $\hat{A}_{\alpha}=0$. Replacing
(\ref{P}) and (\ref{Prepresentation}) into (\ref{L3}) we otain

\begin{align}
\widehat{L} & =\sum_{i,j=1}^{n}m_{ij}^{-1}\hat{p}_{i}\hat{\lambda}'_{j}+\sum_{\alpha,\beta,l,k,j=1}^{n}m_{k\beta}^{-1}\frac{\partial m_{\beta\alpha}}{\partial\widehat{q}_{j}}m_{\alpha l}^{-1}\hat{p}_{l}\hat{p}_{k}\widehat{\pi}'_{j}+\sum_{j=1}^{n}\widehat{Q}_{j}\widehat{\pi}'_{j}+\hat{f}.\label{L4}
\end{align}
Remembering that the function of $\hat{f}$ is to make Hermitian the
Liouvillian (\ref{L4}), we find that the ``quantum'' canonical
transformation (\ref{pQtransformation}) leads to the same Liouvillian
obtained from the KvN theory (see appendix B).

We can proceed a bit further by defining the Hamiltonian operator
\begin{equation}
\hat{\mathscr{H}}=\sum_{i,j=1}^{n}\frac{1}{2}m_{ij}^{-1}\hat{p}_{i}\hat{p}_{j}+\hat{\phi}.\label{hamiltonian}
\end{equation}
The Liouvillian (\ref{L4}) can then be written as

\begin{equation}
\widehat{L}=\sum_{j=1}^{n}\left(\frac{\partial\hat{\mathscr{H}}}{\partial\hat{p}_{j}}\hat{\lambda}'_{j}+\frac{\partial\hat{\mathscr{H}}}{\partial\hat{q}_{j}}\hat{\pi}'_{j}\right)_{+}.\label{L4b}
\end{equation}
The form (\ref{L4b}) concides with the general definition of the
Liouvillian in the KvN theory with the exception of the symmetrization
that is not usually acknowledged (since it is not important for systems
with scalar potentials in Cartesian coordinates).

\section{Action principles}

In this section, we show two methods to obtain the Heisenberg equation
of motion for the position and velocity operators from the variation
of an action operator. The results given here are an expansion of
the ones given in \cite{Schwinger}.

Let us remember here that the time evolution of operator in the Heisenberg
picture is given by

\begin{align}
\hat{q}_{i}(t) & =\hat{U}^{\dagger}(t)\hat{q}_{i}\hat{U}(t),\nonumber \\
\hat{v}_{i}(t) & =\hat{U}^{\dagger}(t)\hat{v}_{i}\hat{U}(t),\label{heisenberg}
\end{align}
where the time evolution operators obeys the Schr\"{o}dinger equation

\begin{equation}
i\frac{\partial}{\partial t}\hat{U}(t)=\hat{L}\hat{U}(t).\label{Sequation}
\end{equation}

We can think of $\hat{q}(t)$ and $\hat{v}(t)$ as defining a path
in operator space, and then we can consider a slightly deformed path
given by

\begin{align}
\hat{q}_{i}(t) & \rightarrow\hat{q}_{i}(t)+\epsilon\eta(t),\nonumber \\
\hat{v}_{i}(t) & \rightarrow\hat{v}_{i}(t)+\epsilon\frac{d\eta(t)}{dt}.\label{variation}
\end{align}
for some differentiable function $\eta(t)$.

We can write the operator version of the Hamiltonian action as

\[
\hat{W}_{H}=\int_{t_{1}}^{t_{2}}\hat{\mathscr{L}}\,dt,
\]
where the Lagrangian operator is given by (\ref{lagrangianoperator}).
In the case where all forces can be derivable from a generalized potential,
the Euler-Lagrange equations in operational form (\ref{eulerlagrange})
follows from demanding that the fixed end-points variation of the
Hamiltonian action vanishes, i.e, after performing the transformation
(\ref{variation}) we impose that $\eta(t_{1})=\eta(t_{2})=0\Rightarrow\delta\hat{W}_{H}=0$.
This result mimics the standard form of the Hamilton principle because
all the operators appearing in $\hat{\mathscr{L}}$ commute with each
other \cite{Schwinger}.

The above, while correct, does not give any new insights compared
to the usual methods of analytical mechanics. On the other hand, the
operational formulation of mechanics allows us to derive the equation
of motion from a different action and a different Lagrangian. Consider
the Schwinger action given by

\begin{align}
\hat{W}_{S} & =\int_{t_{1}}^{t_{2}}\hat{\mathscr{L}_{S}}\,dt,\nonumber \\
\hat{\mathscr{L}_{S}} & =\sum_{i=1}^{n}\left(\frac{d\widehat{q}_{i}}{dt}\hat{\lambda}_{i}^{(q)}+\frac{d\widehat{v}_{i}}{dt}\widehat{\pi}_{i}^{(q)}\right)_{+}-\hat{L}.
\end{align}
Assuming the commutation relations (\ref{comm2}), the Heisenberg
equations of motion (\ref{qv-1}) and (\ref{dv-1}) follow from fixed
end-points variation of the Schwinger action, $\delta\hat{W}_{s}=0$
\cite{Schwinger}. By using the Liouvillian (\ref{L3}) and Eq (\ref{qv-1}),
we can further write the Schwinger Lagrangian as
\begin{equation}
\hat{\mathscr{L}_{S}}=\sum_{i=1}^{n}\left(\left(\frac{d\widehat{v}_{i}}{dt}-\sum_{k=1}^{n}\sum_{j=1}^{3N}\widehat{v}_{k}(\frac{\partial\widehat{X}_{j}}{\partial\widehat{q}_{k}})\frac{\partial\hat{v}_{i}}{\partial\hat{X}_{j}}-\sum_{j=1}^{3N}\frac{1}{m_{j}}\widehat{F}_{j}\frac{\partial\hat{v}_{i}}{\partial\hat{V}_{j}}\right)\widehat{\pi}_{i}^{(q)}\right)_{+}.\label{LS1}
\end{equation}
Notice that $\hat{\mathscr{L}_{S}}$ is a function of non-commuting
operators and that the forces do not need to be derivable from a generalized
potential. 

The Schwinger Lagrangian is not very illuminating when written in
term of the generalized coordinates and velocities. However, the situation
slightly improves when $\mathscr{\hat{L}}_{S}$ is expressed in terms
of Cartesian positions and velocities. The identities (\ref{I1}),
(\ref{I2}), and the inverse transformation 
\begin{align}
\widehat{\pi}_{j} & =\sum_{i=1}^{3N}\frac{\partial\hat{v}_{i}}{\partial\hat{V}_{j}}\widehat{\pi}_{i}^{(q)},
\end{align}
allow us to simplify $\mathscr{\hat{L}}_{S}$ into 
\begin{equation}
\hat{\mathscr{L}_{S}}=\sum_{j=1}^{3N}\left(\left(\frac{d\hat{V}_{j}}{dt}-\frac{1}{m_{j}}\hat{F}_{j}\right)\widehat{\pi}_{j}\right)_{+}.\label{LS2}
\end{equation}
Let us remark that the forces appearing in (\ref{LS2}) are the impressed
ones, the constraining forces are missing. In the absence of constraining
forces, the physical acceleration given by Eq (\ref{VF}) makes $\hat{\mathscr{L}_{S}}$
vanish identically. When there are forces of constraint present, the
Cartesian accelerations $\frac{d}{dt}\widehat{V}_{j}(t)$ are such
that $\hat{W}_{S}$ is stationary.

At the moment, the relation between the Schwinger action principle
and the Hamilton principle is not entirely clear beyond the fact that
both lead to the same equations of motion for the position and velocity
operators. However, it is worth mentioning that the form of the Lagrangian
(\ref{LS2}), and the fact that it deals with acceleration and not
velocities, is quite reminiscent of Gauss' principle of least constraint
\cite{lanczos,gauss1,gauss2,gauss3,gauss4}. Gauss' principle states
that the actual motion of a mechanical system occurring in nature
is such that the accelerations minimize the so-called ``constraint''
function

\begin{equation}
Z=\sum_{i=1}^{3N}\frac{1}{m_{i}}\left(m_{i}a_{i}-F_{i}\right)^{2}.\label{constraintfunction}
\end{equation}
I conjecture that the constraint function (\ref{constraintfunction})
and the Schwinger Lagrangian are closely related, but I can not give
a proof at the moment.

\section{Conclusions}

We have systematically constructed an operational theory for classical
mechanics. We started from the case free of constraint, and later
we incorporated holonomic scleronomous constraints into our formalism.
By taking this unusual approach to describe classical systems, we
have rediscovered several well-known results from analytical dynamics
in an entirely new way.

Our procedure seems to put classical dynamics on new foundations that
are logically independent of the ones invoked in analytical mechanics.
Our main guiding principle was the demand that all transformations
have to be ``quantum'' canonical transformations. Only in a posteriori
analysis can we compare the principles employed and show them to be
(or not) equivalent to each other. For example, Our operational approach
leads to the operational version of the Hamilton principle, while
the usual version of the Hamilton principle leads to Hamiltonian mechanics,
then to the KvN theory, then to our operational approach (see Appendix
B). On the other hand, and it is perhaps the biggest shortcoming of
this paper, we have not given any operational formulation of the D'Alembert
principle. Presently, it is not clear to the author how to even define
the concept of virtual work using operators.

It is also unclear the relationship between the usual Legendre transformation
to obtain Hamiltonian mechanics and the ``quantum'' canonical transformation
we have used to define our momentum representation of the dynamics.
Also, the Hamiltonian (operator) seems to play a rather secondary
role in our approach. We wrote it in Eq (\ref{hamiltonian}) just
for the sake of completeness, but it is not needed at all to describe
the dynamics in terms of the canonical momentum.

The following questions remain open:

(1) The transformation to a momentum description of the dynamics is
given in the constraint-less case by a composition of a scale and
a unitary transformation. Can the same be done for equations (\ref{P})
and (\ref{Prepresentation})?

(2) Is there a relation between the Schwinger action principle and
the Gauss principle of less constraint? How can we justify the Schwinger
principle using only the usual tools from analytical mechanics?

(3) How can we justify our procedure to obtain Lagrange equations
of motion using only analytical tools?

\section*{Appendix A: Spherical coordinates}

Here we give formulas for the ``quantum'' point transformation in
$TQ$ from Cartesian to spherical operators that allow us to write
the Liouvillian accordingly. The transformation is given by

\begin{align}
\widehat{x} & =\widehat{r}\sin\widehat{\theta}\cos\widehat{\phi},\nonumber \\
\widehat{y} & =\widehat{r}\sin\widehat{\theta}\sin\widehat{\phi},\nonumber \\
\widehat{z} & =\widehat{r}\cos\widehat{\theta}.\label{T1}
\end{align}
and 

\begin{align}
\widehat{V}_{x} & =\widehat{V}_{r}\sin\widehat{\theta}\cos\widehat{\phi}+\widehat{r}\widehat{V}_{\theta}\cos\widehat{\theta}\cos\widehat{\phi}-\widehat{r}\widehat{V}_{\phi}\sin\widehat{\theta}\sin\widehat{\phi},\nonumber \\
\widehat{V}_{y} & =\widehat{V}_{r}\sin\widehat{\theta}\sin\widehat{\phi}+\widehat{r}\widehat{V}_{\theta}\cos\widehat{\theta}\sin\widehat{\phi}+\widehat{r}\widehat{V}_{\phi}\sin\widehat{\theta}\cos\widehat{\phi},\nonumber \\
\widehat{V}_{z} & =\widehat{V}_{r}\cos\widehat{\theta}-\widehat{r}\widehat{V}_{\theta}\sin\widehat{\theta}.\label{T2}
\end{align}
The inverse transformation are

\begin{align}
\widehat{r} & =\sqrt{\widehat{x}^{2}+\widehat{y}^{2}+\widehat{z}^{2}},\nonumber \\
\cos\widehat{\theta} & =\frac{\widehat{z}}{\widehat{r}},\qquad\quad\quad\quad\;\:\sin\widehat{\theta}=\frac{\sqrt{\widehat{x}^{2}+\widehat{y}^{2}}}{\widehat{r}},\nonumber \\
\cos\widehat{\phi} & =\frac{\widehat{x}}{\sqrt{\widehat{x}^{2}+\widehat{y}^{2}}},\qquad\sin\widehat{\phi}=\frac{\widehat{y}}{\sqrt{\widehat{x}^{2}+\widehat{y}^{2}}}.
\end{align}

Our task here is to give expressions for the conjugate operators $\hat{\lambda}$
and $\hat{\pi}$. The formulas to be used are the inverse of (\ref{point1})
and (\ref{point2}), for example

\begin{align}
\widehat{\lambda}_{x}= & \left(\frac{\partial\widehat{r}}{\partial\widehat{x}}\widehat{\lambda}_{r}\right)_{+}+\left(\frac{\partial\widehat{\theta}}{\partial\widehat{x}}\widehat{\lambda}_{\theta}\right)_{+}+\left(\frac{\partial\widehat{\phi}}{\partial\widehat{x}}\widehat{\lambda}_{\phi}\right)_{+}\nonumber \\
 & +\left(\frac{\partial\widehat{V}_{r}}{\partial\widehat{x}}\widehat{\pi}_{r}\right)_{+}+\left(\frac{\partial\widehat{V}_{\theta}}{\partial\widehat{x}}\widehat{\pi}_{\theta}\right)_{+}+\left(\frac{\partial\widehat{V}_{\phi}}{\partial\widehat{x}}\widehat{\pi}_{\phi}\right)_{+}.
\end{align}

The results are

\begin{align}
\widehat{\lambda}_{x} & =\sin\widehat{\theta}\cos\widehat{\phi}\,\widehat{\lambda}_{r}+\left(\frac{\cos\widehat{\theta}\cos\widehat{\phi}}{\hat{r}}\widehat{\lambda}_{\theta}\right)_{+}-\left(\frac{\sin\widehat{\phi}}{\hat{r}\sin\widehat{\theta}}\widehat{\lambda}_{\phi}\right)_{+}\nonumber \\
 & +\left(\widehat{V}_{\theta}\cos\widehat{\theta}\cos\widehat{\phi}-\widehat{V}_{\phi}\sin\widehat{\theta}\sin\widehat{\phi}\right)\widehat{\pi}_{r}\nonumber \\
 & +\frac{\sin\widehat{\theta}\cos\widehat{\phi}}{\hat{r}}\left(\widehat{V}_{\theta}\widehat{\pi}_{\theta}\right)_{+}-\left(\frac{\widehat{V}_{\phi}}{\hat{r}}\cos\widehat{\theta}\sin\widehat{\phi}+\frac{\widehat{V}_{r}}{\hat{r}^{2}}\cos\widehat{\theta}\cos\widehat{\phi}\right)\widehat{\pi}_{\theta}\nonumber \\
 & +\frac{\widehat{V}_{r}\sin\widehat{\theta}+\widehat{r}\widehat{V}_{\theta}\cos\widehat{\theta}}{r^{2}\sin^{2}\widehat{\theta}}\widehat{\pi}_{\phi},\nonumber \\
\widehat{\lambda}_{y} & =\sin\widehat{\theta}\sin\widehat{\phi}\,\widehat{\lambda}_{r}+\left(\frac{\cos\widehat{\theta}\sin\widehat{\phi}}{\hat{r}}\widehat{\lambda}_{\theta}\right)_{+}+\left(\frac{\cos\widehat{\phi}}{\hat{r}\sin\widehat{\theta}}\widehat{\lambda}_{\phi}\right)_{+}\nonumber \\
 & +\left(\widehat{V}_{\theta}\cos\widehat{\theta}\sin\widehat{\phi}+\widehat{V}_{\phi}\sin\widehat{\theta}\cos\widehat{\phi}\right)\widehat{\pi}_{r}\nonumber \\
 & +\frac{\sin\widehat{\theta}\sin\widehat{\phi}}{\hat{r}}\left(\widehat{V}_{\theta}\widehat{\pi}_{\theta}\right)_{+}+\left(\frac{\widehat{V}_{\phi}}{\hat{r}}\cos\widehat{\theta}\cos\widehat{\phi}-\frac{\widehat{V}_{r}}{\hat{r}^{2}}\cos\widehat{\theta}\sin\widehat{\phi}\right)\widehat{\pi}_{\theta}\nonumber \\
 & -\frac{\sin\widehat{\phi}}{\hat{r}\sin\widehat{\theta}}\left(\widehat{V}_{\phi}\widehat{\pi}_{\phi}\right)_{+},\nonumber \\
\widehat{\lambda}_{z} & =\cos\widehat{\theta}\,\widehat{\lambda}_{r}-\left(\frac{\sin\widehat{\theta}}{\hat{r}}\widehat{\lambda}_{\theta}\right)_{+}\nonumber \\
 & -\sin\widehat{\theta}\widehat{V}_{\theta}\widehat{\pi}_{r}+\frac{\sin\widehat{\theta}}{\hat{r}^{2}}\widehat{V}_{r}\widehat{\pi}_{\theta}-\frac{1}{\hat{r}}\cos\widehat{\theta}\left(\widehat{V}_{\theta}\widehat{\pi}_{\theta}\right)_{+}.\label{lambda}
\end{align}

and

\begin{align}
\widehat{\pi}_{x} & =\sin\widehat{\theta}\cos\widehat{\phi}\,\widehat{\pi}_{r}+\frac{\cos\widehat{\theta}\cos\widehat{\phi}}{\hat{r}}\widehat{\pi}_{\theta}-\frac{\sin\widehat{\phi}}{\hat{r}\sin\widehat{\theta}}\widehat{\pi}_{\phi},\nonumber \\
\widehat{\pi}_{y} & =\sin\widehat{\theta}\sin\widehat{\phi}\,\widehat{\pi}_{r}+\frac{\cos\widehat{\theta}\sin\widehat{\phi}}{\hat{r}}\widehat{\pi}_{\theta}+\frac{\cos\widehat{\phi}}{\hat{r}\sin\widehat{\theta}}\widehat{\pi}_{\phi},\nonumber \\
\widehat{\pi}_{z} & =\cos\widehat{\theta}\,\widehat{\pi}_{r}-\frac{\sin\widehat{\theta}}{\hat{r}}\widehat{\pi}_{\theta}.\label{pi}
\end{align}

Expressions for the spherical operators can be obtained if desired,
for example

\begin{align}
\widehat{\lambda}_{r} & =\left(\frac{\partial\widehat{x}}{\partial\widehat{r}}\widehat{\lambda}_{x}\right)_{+}+\left(\frac{\partial\widehat{y}}{\partial\widehat{r}}\widehat{\lambda}_{y}\right)_{+}+\left(\frac{\partial\widehat{z}}{\partial\widehat{r}}\widehat{\lambda}_{z}\right)_{+}\nonumber \\
 & +\left(\frac{\partial\widehat{V}_{x}}{\partial\widehat{r}}\widehat{\pi}_{x}\right)_{+}+\left(\frac{\partial\widehat{V}_{y}}{\partial\widehat{r}}\widehat{\pi}_{y}\right)_{+}+\left(\frac{\partial\widehat{V}_{z}}{\partial\widehat{r}}\widehat{\pi}_{z}\right)_{+}\nonumber \\
 & =\left(\sin\widehat{\theta}\cos\widehat{\phi}\widehat{\lambda}_{x}\right)_{+}+\left(\sin\widehat{\theta}\sin\widehat{\phi}\widehat{\lambda}_{y}\right)_{+}+\left(\cos\widehat{\theta}\widehat{\lambda}_{z}\right)_{+}.
\end{align}

\section*{Appendix B: The KvN theory in generalized coordinates}

In this appendix, we investigate the form of the KvN theory of a scleronomous
mechanical system in generalized coordinates.

The KvN theory is a wave mechanics formulation originating from the
Liouville equation of Hamiltonian mechanics,

\begin{equation}
\frac{\partial\rho}{\partial t}=\left\{ H,\rho\right\} ,
\end{equation}
where $\rho=\left|\psi\right|^{2}$ is the probability density, and
$\psi$ is the probability amplitude. The amplitude is a square integrable
function over the whole of the phase space $\varGamma$, this is
\begin{equation}
\int_{\varGamma}\left|\psi\right|^{2}d\omega=1,
\end{equation}
where
\begin{equation}
d\omega=\prod_{j=1}^{n}dq_{j}dp_{j}
\end{equation}
is the Liouville measure in the phase space. Now, the Liouvillian
operator $\hat{L}=-i\left\{ ,H\right\} $ is self-adjoint in the Hilbert
space $L^{2}(\varGamma).$ Due to the linearity in the derivative
of the Poisson brackets, we have that $\psi$ obeys the equation
\begin{equation}
i\frac{\partial\psi}{\partial t}=\hat{L}\psi.
\end{equation}
A Lagrangian of the form
\begin{equation}
\mathscr{L}=\sum_{i,j=1}^{n}\frac{1}{2}m_{ij}v_{i}v_{j}-\phi(q),
\end{equation}
give rise to a Hamiltonian 
\begin{equation}
\mathscr{H}=\sum_{i,j=1}^{n}\frac{1}{2}m_{ij}^{-1}p_{i}p_{j}+\phi(q).
\end{equation}
Hence, we can write the Liouvillian as

\begin{equation}
\hat{L}=-i\sum_{i,j=1}^{n}m_{ij}^{-1}p_{j}\frac{\partial}{\partial q_{i}}-i\sum_{i,j=1}^{n}\frac{1}{2}m_{in}^{-1}\frac{\partial m_{nk}}{\partial q_{j}}m_{k\alpha}^{-1}p_{i}p_{\alpha}\frac{\partial}{\partial p_{j}}+i\sum_{j=1}^{n}\frac{\partial\phi}{\partial q_{j}}\frac{\partial}{\partial p_{j}},
\end{equation}
where the formula of the derivative of an inverse matrix was used.
Defining the self-adjoint operators
\begin{align}
\hat{q}_{i}\psi=q_{i}\psi,\:\; & \;\:\hat{p}_{i}\psi=p{}_{i}\psi,\nonumber \\
\hat{\lambda}'_{j}\psi=-i\frac{\partial\psi}{\partial q_{j}},\;\; & \hat{\pi}'_{j}\psi=-i\frac{\partial\psi}{\partial p_{j}},
\end{align}
and the generalized force $Q_{j}=\frac{\partial\phi}{\partial q_{j}}$,
we have a preliminary form of the Liouvillian in generalized coordinates

\begin{equation}
\hat{L}=\sum_{i,j=1}^{n}m_{ij}^{-1}\hat{p}_{j}\hat{\lambda}'_{i}+\sum_{\alpha,\beta,l,k,j=1}^{n}m_{k\beta}^{-1}\frac{\partial m_{\beta\alpha}}{\partial\widehat{q}_{j}}m_{\alpha l}^{-1}\hat{p}_{l}\hat{p}_{k}\widehat{\pi}'_{j}+\sum_{j=1}^{n}\hat{Q}_{j}\hat{\pi}'_{j}.\label{L5}
\end{equation}
The operator (\ref{L5}) has to be considered just a preliminary form
of the true Liouvillian because it is not Hermitian due to the non-commutativity
of the term $\hat{p}_{i}\hat{p}_{\alpha}\hat{\pi}'_{j}$. The correct
Liouvillian has to be a simmetrized version of (\ref{L5}), but since
there commutator between the $\hat{p}_{i}$ and the $\hat{\pi}'_{j}$
is a c-number, there exist a function $f=f(q,p)$ such that 
\begin{equation}
\hat{L}=\sum_{i,j=1}^{n}m_{ij}^{-1}\hat{p}_{j}\hat{\lambda}'_{i}+\sum_{\alpha,\beta,l,k,j=1}^{n}m_{k\beta}^{-1}\frac{\partial m_{\beta\alpha}}{\partial\widehat{q}_{j}}m_{\alpha l}^{-1}\hat{p}_{l}\hat{p}_{k}\widehat{\pi}'_{j}+\sum_{j=1}^{n}\hat{Q}_{j}\hat{\pi}'_{j}+\hat{f}\label{L6}
\end{equation}
is Hermitian. The operator (\ref{L6}) coincides with (\ref{L4}),
hence, we can conclude that the KvN theory is the wave mechanics version
of the operational theory developed in this paper.

\end{document}